# Monitoring Software Reliability using Statistical Process Control: An MMLE Approach

Dr. R Satya Prasad[1], Bandla Sreenivasa Rao[2] and Dr. R.R. L Kantham3

[1]Department of Computer Science &Engineering, Acharya Nagarjuna University, Guntur, India
profrsp@gmail.com,

[2]Department of Computer Science, VRS & YRN College, Chirala, India
sreenibandla@yahoo.com

3 Department of Statistics, Acharya Nagarjuna University, Guntur, India
kantam_rrl@rediffmail.com

**ABSTRACT**

*This paper consider an MMLE (Modified Maximum Likelihood Estimation) based scheme to estimate software reliability using exponential distribution. The MMLE is one of the generalized frameworks of software reliability models of Non Homogeneous Poisson Processes (NHPPs). The MMLE gives analytical estimators rather than an iterative approximation to estimate the parameters. In this paper we proposed SPC (Statistical Process Control) Charts mechanism to determine the software quality using inter failure times data. The Control charts can be used to measure whether the software process is statistically under control or not.*

**KEYWORDS**

*Software Reliability, Statistical Process Control, Control Charts, NHPP, Exponential Distribution, MMLE, Inter Failure Times Data*

## 1. INTRODUCTION

The software reliability is one of the most significant attributes for measuring software quality. The software reliability can be quantitatively defined as the probability of failure free operation of a software in a specified environment during specified duration.[1]. Thus, probabilistic models are applied to estimate software reliability with the field data. Various NHPP software reliability models are available to estimate the software reliability. The MMLE is one of such NHPP based software reliability model.(2). The software reliability models can be used quantitative management of quality (3). This is achieved by employing SPC techniques to the quality control activities that determines whether a process is stable or not. The objective of SPC is to establish and maintain statistical control over a random process. To achieve this objective, it is necessary to detect assignable causes of variation that contaminate the random process. The SPC had proven useful for detecting assignable causes(4).

## 2. BACKGROUND

This section presents the theory that underlies exponential distribution and maximum likelihood estimation for complete data. If 't' is a continuous random variable with

*pdf*: $f(t; \theta_1, \theta_2, \ldots, \theta_k)$. Where $\theta_1, \theta_2, \ldots, \theta_k$ are $k$ unknown constant parameters which need to be estimated, and *cdf*: $F(t)$. Where, the mathematical relationship between the *pdf* and *cdf* is given by: $f(t) = \frac{d(F(t))}{dt}$. Let 'a' denote the expected number of faults that would be detected given infinite testing time in case of finite failure NHPP models. Then, the mean value function of the finite failure NHPP models can be written as: $m(t) = aF(t)$. where, F(t) is a cumulative distribution function. The failure intensity function $\lambda(t)$ in case of the finite failure NHPP models is given by: $\lambda(t) = aF'(t)$ [5][6]

## 2.1 Exponential NHPP Model

When the data is in the form of inter failure times also called Time between failures, we will try to estimate the parameters of an NHPP model based on exponential distribution [6]. Let N(t) be an NHPP defined as

$$p[N(t)] = \frac{[m(t)]^y}{y!} e^{-m(t)}, y = 0,1,2,\ldots,n$$

Here $m(t)$ is the mean value function of the process of an NHPP given by

$$m(t) = a(1 - e^{-bt}) \quad a>0, b>0, t>=0 \tag{2.1.1}$$

The intensity function of the process is given by

$$\lambda(t) = \frac{dm(t)}{dt} = b(a - m(t)) \tag{2.1.2}$$

## 2.2 Modified Maximum Likelihood Estimation

The constants 'a', 'b' which appear in the mean value function and hence in NHPP, in intensity function (error detection rate) and various other expressions are called parameters of the model. In order to have an assessment of the software reliability 'a',' b' are to be known or they are to be estimated from a software failure data. Suppose we have 'n' time instants at which the first, second, third..., n$^{th}$ failures of a software are experienced. In other words if $S_k$ is the total time to the k$^{th}$ failure, $s_k$ is an observation of random variable $S_k$ and 'n' such failures are successively recorded. The joint probability of such failure time realizations $s_1, s_2, s_3, \ldots s_n$ is

$$L = e^{-m(s_n)} \prod_{k=1}^{n} \lambda(s_k) \tag{2.2.1}$$

The function given in equation (2.1.3)(2.2.1) is called the likelihood function of the given failure data. Values of 'a', ' b' that would maximize L are called maximum likelihood estimators (MLEs) and the method is called maximum likelihood (ML) method of estimation. Accordingly 'a', 'b' would be solutions of the equations

$$\frac{\partial \log L}{\partial a} = 0, \frac{\partial \log L}{\partial b} = 0$$

Substituting the expressions for m(t), λ(t) given by equations (2.1.1) and (2.1.2) in equation (2.2.1), taking logarithms, differentiating with respect to 'a', 'b' and equating to zero, after some joint simplification we get

$$a = \frac{n}{(1-e^{-bn_m})} \tag{2.2.2}$$

$$g(b) = \sum_{k=1}^{n} s_k - \frac{n}{b} + n s_n \frac{e^{-b n_m}}{(1-e^{-b n})} \tag{2.2.3}$$

MLE of 'b' is an iterative solution of equation (2.1.5) (2.2.3) which when substituted in equation (2.1.4) gives MLE of 'a'. In order to get the asymptotic variances and co-variance of the MLEs of 'a', 'b' we needed the elements of the information matrix obtained through the following second order partial derivative.

$$\frac{\partial^2 \log L}{\partial b^2} = \frac{n}{b^2} - n s_n^2 \left[ \frac{1}{(1-e^{-bn_m})} + \frac{e^{-bn_m}}{(1-e^{-bn_m})^2} \right] e^{-bn_m} \tag{2.2.4}$$

Expected values of negatives of the above derivative would be the following information matrix

$$E \begin{bmatrix} -\dfrac{\partial^2 \log L}{\partial a^2} & -\dfrac{\partial^2 \log L}{\partial a \, \partial b} \\ -\dfrac{\partial^2 \log L}{\partial a \, \partial b} & -\dfrac{\partial^2 \log L}{\partial b^2} \end{bmatrix}$$

Inverse of the above matrix is the asymptotic variance covariance matrix of the MLEs of 'a', 'b'. Generally the above partial derivatives evaluated at the MLEs of 'a', 'b' are used to get consistent estimator of the asymptotic variance covariance matrix.

However in order to overcome the numerical iterative way of solving the log likelihood equations and to get analytical estimators rather than iterative, some approximations in estimating the equations can be adopted from [2] [8] and the references there in. We use two such approximations here to get modified MLEs of 'a' and 'b'.
Equation (2.2.3) can be written as

$$\sum_{k=1}^{n} S_k \, a^{n-k} \frac{n}{b} - a \, n \, s_n \, e^{-b n_m} = 0 \tag{2.2.5}$$

Let us approximate the following expressions in the L.H.S of equation (2.2.5) by linear functions in the neighborhoods of the corresponding variables.

$$\frac{s_n \, e^{-b r_n}}{1 - e^{-b r_n}} = m S_n + c, \; n = 1, 2, \ldots n. \tag{2.2.6}$$

where $S_n$ is the slope and 'c' is the intercepts in equations (2.2.6) are to be suitably found. With such values equations (2.2.6) when used in equation (2.2.5) would give an approximate MLE for 'b' as

$$\hat{b} = 1 + \frac{1}{(m S_n + c) + s} \tag{2.2.7}$$

where $s = \sum_{k=1}^{n} \frac{S_k}{n}$

We suggest the following method to get the slopes and intercepts in the R.H.S of equations (2.2.6).

$$F(z) = \frac{n}{n+1} \tag{2.2.8}$$

$$F(Z^*) = p - \sqrt{\frac{pq}{n}} \tag{2.2.9}$$

$$F(Z^n) = p + \sqrt{\frac{pq}{n}} \qquad (2.2.10)$$

Given a natural number 'n' we can get the values of $Z'$ and $Z''$ by inverting the above equations through the function F(z) the L.H.S of equation (2.2.6) we get

$$m = \frac{\left(\frac{z'e^{-z'}}{1-e^{-z'}}\right) - \left(\frac{z''e^{-z''}}{1-e^{-z''}}\right)}{z'-z''} \qquad (2.2.11)$$

$$c = \frac{z'e^{-z'}}{1-e^{-z'}} - m\,z' \qquad (2.2.12)$$

It can be seen that the evaluation of $\xi_n$, C are based on only a specified natural number 'n' and can be computed free from any data. Given the data observations and sample size using these values along with the sample data in equation (2.1.12)(2.2..7) we get an approximate MLE of 'b'. Equation (2.2.2) gives approximate MLE of 'a'.

## 3. ESTIMATION BASED ON INTER FAILURE TIMES DATA

Based on the time between failures data give in Table-1, we compute the software failure process through mean value control chart. We use cumulative time between failures data for software reliability monitoring through SPC. The parameters obtained from Goel-Okumoto model applied on the given time domain data are as follows:

a = 33.396342,
b = 0.003962

'$\hat{a}$' and '$\hat{b}$' are Modified Maximum Likelihood Estimates (MMLEs) of parameters and the values can be computed using analytical method for the given time between failures data shown in Table 1. Using values of 'a' and 'b' we can compute $m(t)$. Now equate the *pdf* of m*(t) to* 0.00135, 0.99865, and 0.5 and the respective control limits are given by

$$T_U = (1 - e^{-bt}) = 0.99865$$
$$T_C = (1 - e^{-bt}) = 0.5$$
$$T_L = (1 - e^{-bt}) = 0.00135$$

These limits are convert at $m(t_u)$, $m(t_c)$ and $m(t_L)$ are given by

$$m(t_u) = 33.3512569382986,\ m(t_c) = 16.6981710073481,\ m(t_L) = 0.04508506100108$$

They are used to find whether the software process is in control or not by placing the points in Mean value chart shown in figure-1. A point below the control limit $m(t_L)$ indicates an alarming signal. A point above the control limit $m(t_u)$ indicates better quality. If the points are falling within the control limits it indicates the software process is in stable [9]. The values of control limits are as shown in Table-2.

**Table-1: Time between failures data** (Xie et al., 2002)

| Failure No. | Time between Failures | Failure No. | Time between Failures | Failure No. | Time between failures | Failure No. | Time between failures | Failure No. | Time between failures |
|---|---|---|---|---|---|---|---|---|---|
| 1 | 30.02 | 7 | 5.15 | 13 | 3.39 | 19 | 1.92 | 25 | 81.07 |
| 2 | 1.44 | 8 | 3.83 | 14 | 9.11 | 20 | 4.13 | 26 | 2.27 |
| 3 | 22.47 | 9 | 21 | 15 | 2.18 | 21 | 70.47 | 27 | 15.63 |
| 4 | 1.36 | 10 | 12.97 | 16 | 15.53 | 22 | 17.07 | 28 | 120.78 |
| 5 | 3.43 | 11 | 0.47 | 17 | 25.72 | 23 | 3.99 | 29 | 30.81 |
| 6 | 13.2 | 12 | 6.23 | 18 | 2.79 | 24 | 176.06 | 30 | 34.19 |

**Table-2: Successive Difference of mean value function**

| Failure No | Cumulative failures | m(t) | m(t) Successive Difference | Failure No | Cumulative failures | m(t) | m(t) Successive Difference |
|---|---|---|---|---|---|---|---|
| 1 | 30.02 | 3.745007495 | 0.168687503 | 16 | 151.78 | 15.09281062 | 1.773292339 |
| 2 | 31.46 | 3.913694999 | 2.511282936 | 17 | 177.5 | 16.86610295 | 0.181718724 |
| 3 | 53.93 | 6.424977934 | 0.1449395 | 18 | 180.29 | 17.04782168 | 0.123892025 |
| 4 | 55.29 | 6.569917434 | 0.362096035 | 19 | 182.21 | 17.1717137 | 0.263324295 |
| 5 | 58.72 | 6.932013469 | 1.348473204 | 20 | 186.34 | 17.435038 | 3.888381284 |
| 6 | 71.92 | 8.280486673 | 0.507278516 | 21 | 256.81 | 21.32341928 | 0.789509245 |
| 7 | 77.07 | 8.787765189 | 0.370602904 | 22 | 273.88 | 22.11292853 | 0.176969998 |
| 8 | 80.9 | 9.158368093 | 1.935032465 | 23 | 277.87 | 22.28989853 | 5.577616276 |
| 9 | 101.9 | 11.09340056 | 1.11713536 | 24 | 453.93 | 27.8675148 | 1.518886819 |
| 10 | 114.87 | 12.21053592 | 0.039414228 | 25 | 535 | 29.38640162 | 0.03590267 |
| 11 | 115.34 | 12.24995015 | 0.515572704 | 26 | 537.27 | 29.42230429 | 0.238631489 |
| 12 | 121.57 | 12.76552285 | 0.275243684 | 27 | 552.9 | 29.66093578 | 1.420599455 |
| 13 | 124.96 | 13.04076653 | 0.72160932 | 28 | 673.68 | 31.08153524 | 0.266001157 |
| 14 | 134.07 | 13.76237585 | 0.168851459 | 29 | 704.49 | 31.34753639 | 0.259556189 |
| 15 | 136.25 | 13.93122731 | 1.161583304 | 30 | 738.68 | 31.60709258 | |

### 4. CONTROL CHART

Control charts are sophisticated statistical data analysis tools, which include upper and lower limits to detect any outliers. They are frequently used in SPC analysis [10]. We used control chart mechanism to identify the process variation by placing the successive difference of cumulative mean values shown in table 2 on y axis and failure number on x axis and the values of control limits at mean value function are placed on Inter Failure Control chart, we obtained Figure 1. The Inter Failure Control chart shows that the successive differences of *m(t)* at 10$^{th}$ and 25$^{th}$ failure data has fallen below $m(t_L)$, which indicates the failure process is identified. It is significantly early detection of failures using Inter Failure Control chart. The software quality is determined by detecting failures at an early stage. The remaining failure data shown in Figure-1 is stable. No failure data fall outside $m(t_u)$. It does not indicate any alarm signal.

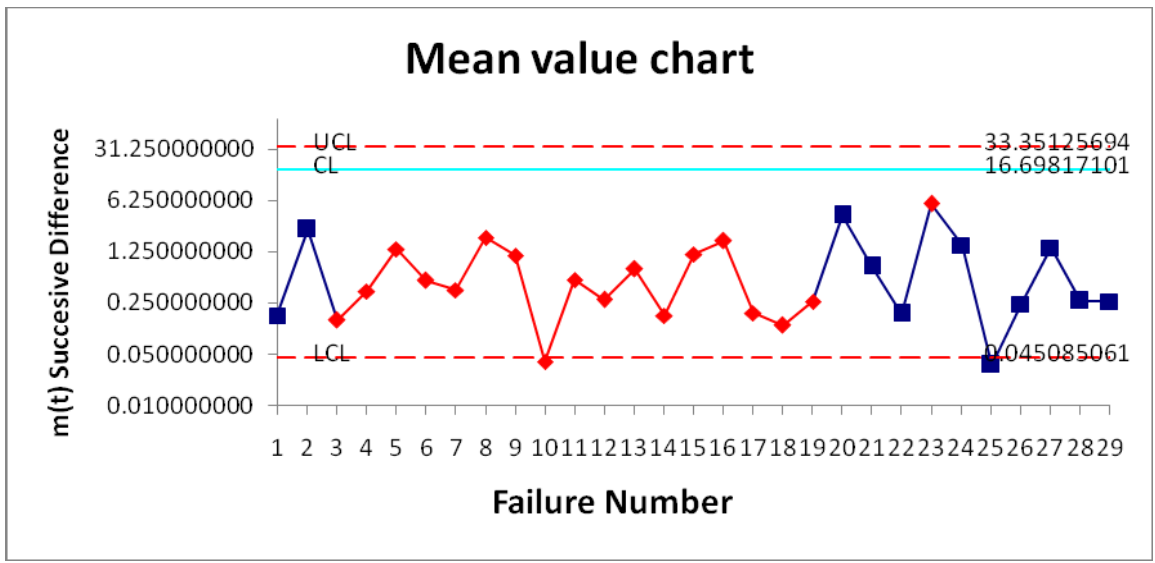

## 5. CONCLUSION

This Mean value chart (Fig 1) exemplifies that, the first out – of – control and second our-of-control situation is noticed at the 10th failure and 25th failure with the corresponding successive difference of m(t) falling below the LCL. It results in an earlier and hence preferable out - of - control for the product. The assignable cause for this is to be investigated and promoted. The out of control signals in and the model suggested in Satya Prasad at el [2011] [ 13 ] are the same. We therefore conclude that adopting a modification to the likelihood method doesn't alter the situation, but simplified the procedure of getting the estimates of the parameters, thus resulting in a preference of the present model to the one described in Satya Prasad et al [2011] [ 13 ].

## 6. ACKNOWLEDGMENT

We gratefully acknowledge the support of Department of Computer Science and Engineering, Acharya Nagarjuna University and PG Department of Computer Applications, VRS & YRN College, Chirala for providing necessary facilities to carry out the research work.

## 7. REFERENCES

1. Musa, J.D, Software Reliability Engineering McGraw-Hill, 1998
2. Kantam, R.R.L;and Dharmarao V (1994) "Half Logistic Distribution –An improvement over M.L Estimator",proceedings of 11 Annual conference of SDS, 39-44.
3. Quantitative quality management through defect prediction and statistical process control (2nd World Quality Congress for Software, Japan, September 2000.).
4. *Florac*, *W.A.*, *Carleton*, *A.D.*, "Measuring The Software Process:" Addison-wesley Professional, Jul 1999.
5. Pham. H., 2003. "Handbook of Reliability Engineering", Springer.
6. Pham. H., 2006. "System software reliability", Springer.
7. Kantam, R.R.L;and Sriram B (2001) "Variable Control Charts Based on Gama Distribution", IAPQR Transactions 26(2), 63-78.
8. MacGregor, J.F., Kourti, T., 1995. "Statistical process control of multivariate processes". Control Engineering Practice Volume 3, Issue 3, March 1995, Pages 403-414 .
9. Koutras, M.V., Bersimis, S., Maravelakis,P.E., 2007. "Statistical process control using shewart control charts with supplementary Runs rules" Springer Science Business media 9:207-224.

### AUTHORS PROFILE

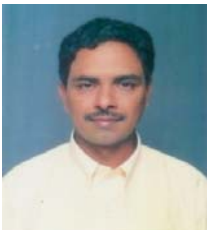

Dr. R. Satya Prasad received Ph.D. degree in Computer Science in the faculty of Engineering in 2007 from Acharya Nagarjuna University, Andhra Pradesh. He received gold medal from Acharya Nagarjuna University for his out standing performance in Masters Degree. He is currently working as Associate Professor and H.O.D, in the Department of Computer Science & Engineering, Acharya Nagarjuna University. His current research is focused on Software Engineering, Software reliability. He has published several papers in National & International Journals.

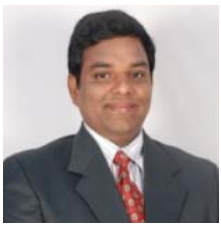

Bandla Srinivasa rao received the Master Degree in Computer Science and Engineering from Dr MGR Deemed University, Chennai, Tamil Nadu, India.

He is Currently working as Associate Professor in PG Department of Computer Applications, VRS & YRN College, Chirala, Andhra Pradesh, India. His research interests include software reliability, Cryptography and Computer Networks. He has published several papers in National and International Journals.

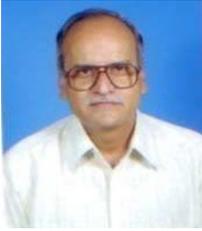 R.R.L Kantam is a professor of statistics at Acharya Nagarjuan University, Guntur,India. He has 31 years of teaching experience in statistics at Under Graduate and Post Graduate Programs. As researcher in statistics, he has successfully guided 8 students for M.Phil in statistics and 5 students for Ph.D. in statistics. He has authored 54 research publications appeared various statistical journals published in India and other countries Like US, UK, Germany, Pakistan, Srilanka, and Bangladesh. He has been a referee for Journal of Applied Statistics (UK), METRON (Italy), Pakistan Journal of Statistics ( Pakistan), IAPQR- Transactions (India), Assam Statistical Review (India) and Gujarat Statistical Review( India). He has been a special speaker in technical sessions of a number of seminars and Conferences, His area of research interest are Statistical Inference, Reliability Studies, Quality Control Methods and Actuarial Statistics. As a teacher his present teaching areas are Probability Theory, Reliability, and Actuarial Statistics. His earlier teaching topics include Statistical Inference, Mathematical analysis, Operations Research, Econometrics, Statistical Quality Control, Measure theory.